\pdfoutput=1
\documentclass[doublecol]{epl2}
\usepackage{amsmath,amssymb}
\usepackage{xcolor}
\usepackage{graphicx}
\usepackage[colorlinks,linkcolor=blue,urlcolor=blue,citecolor=blue]{hyperref}
\usepackage{xspace}

\newcommand{\eref}[1]{eq.~(\ref{#1})}%
\newcommand{\erefss}[1]{eqs.~(\ref{#1})}%
\newcommand{\Eref}[1]{Equation~(\ref{#1})}%
\newcommand{\fref}[1]{fig.~\ref{#1}} %

\newcommand{\St}{\Delta S_\mathrm{tot}}
\newcommand{\Sm}{\Delta S_\mathrm{med}}
\newcommand{\dSs}{S_\mathrm{sys}}
\newcommand{\Ss}{\Delta S_\mathrm{sys}}
\newcommand{\tSt}{\Delta \tilde{S}_\mathrm{tot}^\mathrm{A}}
\newcommand{\tSm}{\Delta \tilde{S}_\mathrm{med}^\mathrm{A}}
\newcommand{\tSs}{\Delta \tilde{S}_\mathrm{sys}^\mathrm{A}}

\begin{document}

\title{Fluctuation theorem for entropy production of a partial system
in the weak coupling limit}

\author{Deepak Gupta and Sanjib Sabhapandit}
\institute{Raman Research Institute - Bangalore 560080, India}
\date{\today}

{Nonequilibrium and irreversible thermodynamics}
{Fluctuation phenomena, random processes, noise, and Brownian motion}

\abstract{
Small systems in contact with a heat bath evolve by stochastic
dynamics.  Here we show that, when one such small system is weakly
coupled to another one, it is possible to infer the presence of such
weak coupling by observing the violation of the steady state
fluctuation theorem for the partial entropy production of the observed
system.  We give a general mechanism due to which the violation of the
fluctuation theorem can be significant, even for weak coupling.  We
analytically demonstrate on a realistic model system that this
mechanism can be realized by applying an external random force to the
system. In other words, we find a new fluctuation theorem for the
entropy production of a partial system, in the limit of weak coupling.
}

\maketitle

\textbf{Introduction.} -- 
In systems where a few slow degrees of freedom (e.g., those of a
colloidal particle in water) interact with a large number of fast
degrees of freedom (e.g., those of the water molecules) ---and there
is a clear separation of timescales between the fast and slow degrees
of freedom--- the effects of the fast degrees of freedom on the slow
degrees of freedom can be replaced by an effective white noise (and
dissipation). This leads to a stochastic dynamics for the slow degrees
of freedom where the fast degrees of freedom act as a heat bath.
Within the framework of stochastic thermodynamics, the heat exchange
between a stochastic system (of slow degrees of freedom) and a bath,
and the work done on a stochastic system can be defined along
individual stochastic trajectories~\cite{sekimoto2010, jarzynski2011,
sevick2008, Campisi:2011, seifert2012, Van-den-Broeck:2015,
Esposito:2009}.  While fluctuations of thermodynamic variables usually
do not play any role in macroscopic systems, they are important for
small systems~\cite{bustamante2005, verley2014, martinez2015,
blickle2011, zon2004a, tietz2006, wang2002, Carberry:2004aa,
liphardt2002, collin2005, Koski:2013aa, Ciliberto:2013aa} consisting
of a few slow degrees of freedom, where the energies are comparable to
$k_B T$.  Over the past two decades or so, a number of remarkable
mathematical relations have been found concerning the fluctuations of
entropy~\cite{evans1993, gallavotti1995, kurchan1998, lebowitz1999,
seifert2005}, work~\cite{jarzynski1997, crooks1999}, and
heat~\cite{zon2003, Saito:2007, Speck:2005, Jarzynski:2004, Noh:2012}.

The stochastic entropy production in a bath (medium) with a
temperature $T$, due to an amount of heat $Q$ extracted from it by a
stochastic system, is given by $\Sm=-Q/T$. By assigning a certain
entropy $\dSs$ for the stochastic system along the
trajectories~\cite{seifert2005}, the total entropy production is
defined as $\St=\Sm+\Ss$. In equilibrium $\St=0$, whereas
non-equilibrium processes generate entropy.  The fluctuation theorem
(FT) relates the probability density functions (PDFs) of positive and
negative entropy productions in a given duration $t$, in the steady state
of a non-equilibrium process, by~\cite{evans1993, gallavotti1995,
kurchan1998, lebowitz1999, seifert2005}.
\begin{equation}
\ln \bigl[P_t(\St)/P_t(-\St)\bigr]=\St, 
\label{FT}
\end{equation}
where the Boltzmann's constant is set to unity ($k_B=1$).


\begin{figure}
\centerline{\includegraphics[width=.8\hsize]{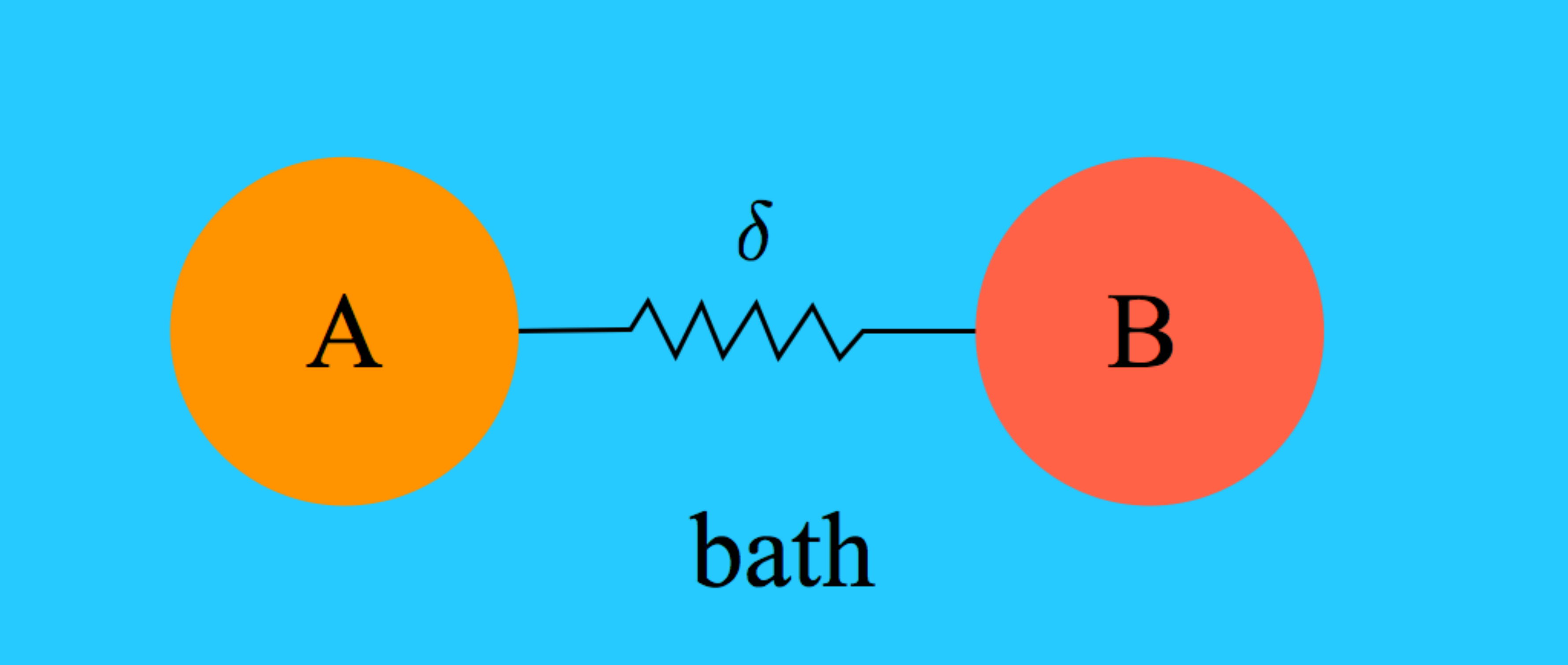}}
\caption{\label{schematic} (Color online) Schematic diagram of  two stochastic
systems A and B coupled to each other by a dimensionless coupling
strength $\delta$. Both are in contact with the same heat bath at a
temperature $T$.}
\end{figure}


Now consider a non-equilibrium system A coupled to another stochastic
system B, with a dimensionless coupling strength $\delta$
(see \fref{schematic}). The total entropy production of the combined
systems A and B would evidently satisfy the FT, given by \eref{FT}. On
the other hand, the FT is not expected to hold, if the total entropy
production is measured by considering only system A~\cite{Mehl:2012fw,
Pietzonka:2014bf, Borrelli:2015eq, Ribezzi-Crivellari:2014, Chun2015,
Rahav:2007aa, Puglisi:2010aa, Shiraishi:2015aa, Talkner:2009aa}.
Therefore, any deviations from the FT as given by \eref{FT} would
infer coupling to other stochastic processes.  Clearly, without any
coupling between A and B, the total entropy production for system A
would again satisfy the FT.  The question thus naturally arises
whether there is any deviation from the FT, given by \eref{FT}, for
vanishingly small coupling ($\delta\to 0$).

In this letter, we show that under certain driving protocols, there
can be significant deviations from \eref{FT}, even in the limit
$\delta\to 0$, yielding a new FT for the entropy production of a
partial system. This finding is in contrast with earlier
studies~\cite{Mehl:2012fw, Pietzonka:2014bf, Borrelli:2015eq,
Ribezzi-Crivellari:2014, Chun2015, Rahav:2007aa, Puglisi:2010aa,
Shiraishi:2015aa, Talkner:2009aa}, where the order of violation of the
FT scales with the coupling strength which smoothly disappears in the
limit of coupling strength going to zero. We propose a general
mechanism, by which the FT for the partial entropy production can be
broken under weak coupling. We demonstrate this mechanism for an
experimentally realizable prototypical system.

\textbf{General mechanism.} --
For a Markov process, to obtain the PDF $P_t(\Delta S)$ of the
entropy production $\Delta S$ (or heat, work, etc.) in the steady
state, in a given time $t$, usually, one first needs consider the
joint PDF $P_t(\Delta S, U)$ of $\Delta S$ and all the relevant
stochastic (slow) variables (denoted by the set $U$) that describe the
system. The joint PDF satisfies a Fokker-Planck equation (FPE),
$[\tau\partial_t - \mathcal{L}] P_t(\Delta S, U)=0$, with the initial
condition $P_0(\Delta S, U) =\delta (\Delta S) \delta (U-U_0)$. Here,
$\tau$ is the characteristic time of the system. The Fokker-Planck
operator $\mathcal{L}$ involves differential operators with respect to
$U$ as well as $\Delta S$ and the exact form of $\mathcal{L}$ depends
on the Langevin equations that describe the system. Integrating out
$\Delta S$ from the joint PDF yields the PDF of the stochastic
variables $P_t(U)=\int P_t(\Delta S, U)\, d\Delta S$ and in the limit
$t\to\infty$ we get the steady state PDF $P_{t\to\infty}(U) \to
P_\mathrm{ss}(U)$. The PDF of the entropy production $P_t(\Delta S)$ can be obtained from joint distribution $P_t(\Delta S,U)$ by integrating out $U$, and averaging over the initial
variables $U_0$ with respect to the steady state PDF
$P_\mathrm{ss}(U_0)$.  It turns out that, the FT for the total entropy
production $\Delta S_\mathrm{tot}$ of a complete stochastic system, as
given by \eref{FT}, can be proven~\cite{seifert2005} without the
explicit form of $P_t(\Delta S_\mathrm{tot})$, and hence, the proof
does not require solving the FPE.  However, this is not the case for
other quantities such as heat, work, or the entropy production of a
partial system ---which is the observable of interest of this
letter. For these quantities, there is no general proof for a FT and
in some cases, the FT (for those quantities) is not even satisfied.
Therefore, for work, heat, entropy production of partial system, etc.,
one has to rely on the explicit form of the PDFs to make any
statement.  Unfortunately, in practice, finding the solution of the
FPE is a non-trivial task and there exists only a few examples where
the complete solution of the FPE can be obtained.  Therefore, the next
best thing is to try to obtain the PDFs for large $t$, and examine
whether the FT is satisfied at least for large $t$. This is our goal
in this letter.

Now let $U_\mathrm{A}\subset U$ be the set of stochastic variables
that describe system A.  The PDF $P_t^\mathrm{A}(U_\mathrm{A})$ at any
time $t$ can be found by keeping only the degrees of freedom of
sub-system A and integrating out the rest from $P_t(U)$. The system
entropy for A can be defined as~\cite{seifert2005}
\begin{math}
\dSs^\mathrm{A}(t) =-\ln
P_t^\mathrm{A}(U_\mathrm{A}(t)).
\end{math}
Let $\Sm^\mathrm{A}=-Q_\mathrm{A}/T$ be the entropy production in the
medium due to A, where $Q_\mathrm{A}$ is the heat transfer from the
medium to A in a given duration. The joint PDF $P_t(\Sm^\mathrm{A},
U)$ satisfies a FPE as mentioned above.  

It is convenient to consider the generating function
\begin{math}
Z(\lambda, U,t|U_0)=\bigl\langle \exp(-\lambda \Sm^\mathrm{A}) \bigr\rangle_{(U,U_0)}, 
\end{math}
where the expectation is taken over all trajectories of the system
that evolve from a given initial configuration $U_0$ to a given final
configuration $U$ in a given duration $t$. Clearly, $Z(0, U,
t|U_0)=P_t(U)$ with the initial condition $P_0(U)=\delta(U-U_0)$.  The
FPE for $P_t(\Sm^\mathrm{A}, U)$ would lead to a Fokker-Planck-like
equation $[\tau\partial_t -\mathcal{L}_\lambda]Z(\lambda, U, t|U_0)=0$
with the initial condition $Z(\lambda, U, 0|U_0)=\delta(U-U_0)$.  The
differential operator $\mathcal{L}_\lambda$ reduces to $\mathcal{L}$
for $\lambda=0$. The solution for $Z(\lambda, U, t| U_0)$ can be
expressed in the eigenbasis of the operator $\mathcal{L}_\lambda$ as
\begin{equation*}
Z(\lambda, U, t|U_0)=\sum_n \chi_n(U_0,\lambda) \Psi_n(U,\lambda)
e^{(t/\tau)\mu_n(\lambda)}.
\end{equation*}
Here $\{\mu_n(\lambda)\}$ are the eigenvalues of $\mathcal{L}_\lambda$
and $\{\chi_n(U,\lambda)\}$ and $\{\Psi_n(U,\lambda)\}$ are the left
and right eigenfunctions, which satisfy the eigenvalue equation 
$\mathcal{L}_\lambda \Psi_n(U,\lambda) = \mu_n(\lambda)
\Psi_n (U,\lambda)$ and the orthonormality  $\int \chi_m(U,\lambda)
\Psi_n(U,\lambda)\, dU=\delta_{m,n}$. 
The large time behavior is determined by the term containing the
largest eigenvalue. Let $\mu(\lambda):=\max\{\mu_n(\lambda)\}$ be the
largest eigenvalue and $\chi(U,\lambda)$ and $\Psi(U,\lambda)$,
respectively, be the corresponding eigenfunctions. Thus for large time,
\begin{equation}
Z(\lambda, U, t|U_0) =\chi(U_0,\lambda) \Psi(U,\lambda)
e^{(t/\tau)\mu(\lambda)} + \dotsb.
\label{z0}
\end{equation}
Evidently, $\mu(0)=0$, $\chi(U_0,0)=1$, and
$P_\mathrm{ss}(U)=P_{t\to\infty}(U)=\Psi(U,0)$ is the steady-state PDF
of $U$. Consequently, the steady-state PDF $P_\mathrm{A}^\mathrm{ss}
(U_\mathrm{A})= P_{t\to\infty}^\mathrm{A} (U_\mathrm{A})$ can be
obtained from $\Psi(U,0)$.

In the steady state, the change in the system-entropy of A in duration
$t$, is given by $\Ss^\mathrm{A}=\ln \bigl[P_\mathrm{A}^\mathrm{ss}
(U_{0\mathrm{A}})/P_\mathrm{A}^\mathrm{ss} (U_\mathrm{A})\bigr] $.
Therefore, the generating function of the total entropy production,
$\St^\mathrm{A}=\Sm^\mathrm{A}+\Ss^\mathrm{A}$ of A, in the steady
state can be obtained, using \eref{z0} in $Z(\lambda,
U,t|U_0) \exp(-\lambda \Ss^\mathrm{A})$ and averaging over the initial
condition $U_0$ with respect to the steady-state PDF $\Psi(U_0,0)$ and
integrating over the final variables $U$, as
\begin{equation}
\bigl\langle \exp(-\lambda \St^\mathrm{A}) \bigr\rangle =
g(\lambda) e^{(t/\tau)\mu(\lambda)} +\dotsb, 
\label{Z}
\end{equation}
where
\begin{align*}
g(\lambda)=\int dU_0\Psi(U_0,0)
\bigl[P_\mathrm{A}^\mathrm{ss}
(U_{0\mathrm{A}})]^{-\lambda} \chi(U_0,\lambda) 
\notag\\ \times
\int dU
\bigl[P_\mathrm{A}^\mathrm{ss}
(U_\mathrm{A})\bigr]^{\lambda} \Psi(U,\lambda). 
\end{align*}

The PDF is related to the above generating function, by the inverse
transformation
\begin{equation*}
P(\St^\mathrm{A})
=  \frac{1}{2\pi i} \int_{-i\infty}^{+i \infty} 
\bigl\langle \exp(-\lambda \St^\mathrm{A}) \bigr\rangle\, e^{\lambda
\St^\mathrm{A}} \, d\lambda. 
\end{equation*}
Therefore, for large $t$, the PDF of the time-averaged total entropy
production $s=(t/\tau)^{-1}\St^\mathrm{A}$ is given by
\begin{equation}
p(s)
= \frac{(t/\tau)}{2\pi i} \int_{-i\infty}^{+i \infty}
g(\lambda)\, e^{(t/\tau)[\mu(\lambda) + \lambda s]}\, d\lambda +\dotsb. 
\label{PDF}
\end{equation}

First consider the case in which system A is isolated from other
stochastic systems ($\delta=0$). Here we use the notations
$g_0(\lambda)$ and $\mu_0(\lambda)$ in place of $g(\lambda)$ and
$\mu(\lambda)$ respectively.  For this isolated system, the FT as
in \eref{FT} must hold for $\St^\mathrm{A}$, that is,
$p(s)/p(-s)=\exp[(t/\tau) s]$. From \eref{PDF}, with the change of the
integration variable $\lambda\to 1-\lambda$, we get
\begin{align*}
e^{(t/\tau) s} p(-s)=
\frac{(t/\tau)}{2\pi i} \int_{1-i\infty}^{1+i \infty}
g_0(1-\lambda)\, e^{(t/\tau)[\mu_0(1-\lambda) + \lambda s]}\,
d\lambda \notag\\
+\dotsb. 
\end{align*}
Note that the contour of integration C in the above integral is
parallel to the imaginary axis through $\text{real}(\lambda)=1$.  Now,
for the right hand side to be equal to $p(s)$, we require the
Gallavotti-Cohen (GC) symmetry $\mu_0(\lambda)=\mu_0(1-\lambda)$ and
$g_0(\lambda)=g_0(1-\lambda)$ to hold, and both $\mu_0(\lambda)$ and
$g_0(\lambda)$ to be analytic, at least within the region between the
imaginary axis and and the contour C, so that the contour C can be
shifted to the imaginary axis through the origin without any
additional contribution from singularities. For $t \gg\tau$, the
saddle-point approximation of \eref{PDF} gives
\begin{align*}
p(s) =\sqrt{\frac{(t/\tau)}{2\pi |\mu''_0(\lambda_0^*)|}}\, g_0(\lambda_0^*)\,
e^{(t/\tau)[\mu_0(\lambda_0^*) + \lambda_0^* s]} +
O\left(\sqrt{\frac{\tau}{t}}\right),
\label{saddle-point approximation}
\end{align*}
where the saddle point $\lambda_0^*(s)$ is given by 
\begin{equation}
\mu_0'(\lambda_0^*)=-s.
\label{saddle-point condition}
\end{equation}
 The saddle point satisfies the symmetry $\lambda_0^*(s)
+\lambda_0^*(-s)=1$.  Now, ignoring the subleading prefactor, one gets
the so-called large deviation form~\cite{touchette2009}:
$p(s) \sim \exp\bigl[(t/\tau) I_0(s)\bigr]$, where the function
$I_0(s)$ is usually known as the large deviation function (LDF), given
by $I_0(s)=\mu_0(\lambda_0^*) +\lambda_0^* s$.  The GC symmetry
implies the symmetry
\begin{equation}
I_0(s)-I_0(-s)=s, 
\label{LDF symmetry}
\end{equation}
which is equivalent to the FT as in \eref{FT} for large
$t$. In this letter, our aim is to investigate, whether such
a relation is valid for non-zero $\delta$, in the limit $\delta\to 0$.

Let us consider the situation where $\mu_0(\lambda)$ is analytic only
within a finite region bounded by a pair of branch point singularities
at $\lambda_\pm$. 
For the FT to hold, $g_0(\lambda)$ must be analytic  within
this region $\lambda \in (\lambda_-, \lambda_+)$, with $\lambda_-
<0$ and $\lambda_+>1$ with $\lambda_++\lambda_-$=$1$. Moreover,
$\mu_0(\lambda_+)=\mu_0(1-\lambda_+)=\mu_0(\lambda_-)$. We assume that
near these branch points\footnote{\label{note1}This specific branch point behavior is taken only as an explicit example. It is not necessary to have this specific form and
 one can in fact have other branch point behaviors such as the
 logarithmic one. The nature of the branch point singularity is note important as it only contributes to the subleading correction.}
\begin{equation*}
\mu_0(\lambda)=
\begin{cases}
\mu_0(\lambda_+) - b (\lambda_+-\lambda)^{\rho_0} + \dotsb
 & \text{as $\lambda\to \lambda_+$},  \\
\mu_0(\lambda_-) - b (\lambda-\lambda_-)^{\rho_0} +\dotsb
 & \text{as $\lambda\to \lambda_-$} ,
\end{cases}
\label{mu_0 near BP}
\end{equation*}
where $0< \rho_0 <1$, and $b$ is a constant. From the saddle-point
equation~(\ref{saddle-point condition}), it follows that
$\lambda_0^*(s) \to \lambda_\pm$ at the leading order as
$s\to \mp\infty$.  Consequently $I_0(s) = \mu_0(\lambda_\pm)
+ \lambda_\pm s +\dotsb$ at the leading order in $s$, as
$s\to \mp\infty$.  Since \eref{LDF symmetry} is valid for all $s$, the
subleading correction terms to the relation $I_0(s) -
I_0(-s)=[\lambda_+ + \lambda_-] s$ vanish at all orders.

In the presence of a non-zero coupling ($\delta>0$), let us
suppose that $\mu(\lambda)$ has branch points at
$\lambda_\pm^{(\delta)}$. The saddle-point approximation of \eref{PDF}
gives the large deviation form $p(s) \sim \exp\bigl[(t/\tau)
I(s)\bigr]$.
If $g(\lambda)$ is analytic in the region
 $\lambda\in\bigl(\lambda_-^{(\delta)}, \lambda_+^{(\delta)}\bigr)$,
 then the LDF is given by $I(s)=\mu(\lambda^*) +\lambda^* s$ with
 $\mu'(\lambda^*)=-s$, as in the $\delta=0$ case.  In case
 $g(\lambda)$ has a singularity within this range, it can change the
 LDF~\cite{sabhapandit2011, sabhapandit2012, pal2013}. However,
 eventually, we are interested in the $\delta\to0$ limit, where we can
 write $g(\lambda)=g_0(\lambda) + \delta^{c} g_1(\lambda)+\dotsb$,
 with $c>0$, and the function $g_1(\lambda)$ may have
 singularities. Therefore, the integral in \eref{PDF} can be written
 as the sum of two integrals, one with a prefactor proportional to
 $g_0(\lambda)$ and the other with the prefactor proportional to
 $\delta^c g_1(\lambda)$. It is evident that the second integral would
 not contribute in the $\delta\to0$ limit.  Therefore, we only
 consider the integral with the prefactor proportional to
 $g_0(\lambda)$. As in the $\delta=0$ case, here we get $I(s)
 = \mu\bigl(\lambda_\pm^{(\delta)}\bigr) + \lambda_\pm^{(\delta)} s
 +\dotsb$ as $s \to \mp\infty$, at the leading order in $s$, where
 $\mu\bigl(\lambda_\pm^{(\delta)}\bigr)$ represents the analytical part
 of $\mu(\lambda)$ at the respective branch points. This implies that
 the asymmetry function,
\begin{equation}
f(s)=I(s)-I(-s), 
\label{asymmetry function}
\end{equation}
has the asymptotic form:
\begin{equation*}
f(s)
= \bigl[\mu\bigl(\lambda_-^{(\delta)}\bigr)
- \mu\bigl(\lambda_+^{(\delta)}\bigr) \bigr]+\bigl[\lambda_+^{(\delta)}
+ \lambda_-^{(\delta)}\bigr] s+\dotsb~\text{as}~s\to\infty.
\end{equation*}
Since, we do not expect $\mu(\lambda)$ to obey the GC symmetry, the
slope $\bigl[\lambda_+^{(\delta)} + \lambda_-^{(\delta)}\bigr]$ need
not be unity and $f(s)$ can have subleading corrections in
$s$. Therefore, for a finite $\delta$, the deviations from the
straight line $f(s)=s$, provides a measure of the violation of the
FT. Now the question is whether such deviation persists in the limit
$\delta\to 0$.

To answer this question, we note that, purely on the general ground,
in the limit $\delta\to 0$, there can be four distinct possibilities:
\begin{equation}
\label{possibilities}
\begin{split}
\text{(P1)}~& \lambda_\pm^{(\delta)}\to \lambda_\pm,\\
\text{(P2)}~& \lambda_+^{(\delta)}\to \lambda_+
~~\text{and}~~ \lambda_-^{(\delta)}\to \tilde\lambda_-, \\
\text{(P3)}~&
\lambda_-^{(\delta)}\to \lambda_- ~~\text{and}~~
\lambda_+^{(\delta)}\to \tilde\lambda_+, \\
\text{(P4)}~& \lambda_\pm^{(\delta)}\to \tilde\lambda_\pm. 
\end{split}
\end{equation}
Note that the original contour of integration
is along the imaginary axis through the origin.  Therefore, the two
singularities, one on each side of the origin, that are closest to the
origin from the respective side only matter, as the saddle point is
bounded by these two closest singularities.  Hence,
$\lambda_-< \tilde\lambda_-<0$ and $0 < \tilde\lambda _+ <\lambda_+$.

Clearly, the FT is not violated for the case (P1) in \eref{possibilities} and one gets the same LDF $I_0(s)$ obtained above for the uncoupled case.  Now, consider the situation (P2). In this case, near $\tilde\lambda_-$, in the limit $\delta\to 0$ we can
write (see footnote~\textsuperscript{\ref{note1}}):
\begin{math}
\mu(\lambda)=\mu_0 (\lambda)
- a \delta^\beta (\lambda-\tilde\lambda_-)^\rho,
\end{math}
where  $\beta>0$, $0<\rho<1$, and $a$ is a constant. The
saddle-point equation $\mu'(\lambda^*)=-s$, yields
\begin{equation}
\mu'_0(\lambda^*) - \frac{a\,\rho \,\delta^\beta}{(\lambda^*
-\tilde\lambda_-)^{1-\rho}} = -s
\label{SPE1}
\end{equation}
We have found above for the $\delta=0$ case that the saddle point given by \eref{saddle-point condition} stays between $(\lambda_-, \lambda_+)$. Therefore, for $\delta\to 0$, if
$\lambda_-^{(\delta)}\to \tilde\lambda_-$ instead of $\lambda_-$, then
it is necessary that $\lambda_- < \tilde\lambda_- <0$. In the limit
$\delta\to 0$, when $s$ increases from $-\infty$ to $\infty$, the
saddle point $\lambda^*(s)$ moves from $\lambda_+$ to
$\tilde\lambda_-$ on the real $\lambda$ line. It is evident
from \eref{SPE1} that for $[\lambda^*(s)-\tilde\lambda_-
]\gg \delta^{\beta/(1-\rho)}$, the left hand side of \eref{SPE1} is
dominated by the first term and therefore the saddle-point is given by
the equation $\mu'_0(\lambda^*) = -s$.  Consequently, the LDF is the
same $I_0(s)$, that has been obtained for the uncoupled case. On the
other hand, for
$[\lambda^*(s)-\tilde\lambda_-] \ll \delta^{\beta/(1-\rho)}$, the
second term on the left hand side dominates the first term, which
results in
\begin{math}
\lambda^*(s)=\tilde\lambda_- +
O\bigl[(\delta^\beta/s)^{1/(1-\rho)}\bigr]. 
\end{math}
This gives $I(s)=\mu_0(\tilde\lambda_-) +
\tilde\lambda_- s +
O\bigl[(\delta^\beta/s^\rho)^{1/(1-\rho)}\bigr]. $ Thus, in the
limit $\delta\to0$, we get
\begin{equation}
I(s)=
\begin{cases}
I_0(s)  &\text{for $s<s_1^*$}\\
\mu_0(\tilde\lambda_-) + \tilde\lambda_- s &\text{for $s>s_1^*$}
\end{cases}
\end{equation}
where $s_1^*$ is given by $\lambda_0^*(s_1^*)=\tilde\lambda_-$.
Similarly for (P3) in \eref{possibilities}, we get
\begin{equation}
I(s)=
\begin{cases}
\mu_0(\tilde\lambda_+) + \tilde\lambda_+ s &\text{for $s<s_2^*$}\\[1mm]
I_0(s)  &\text{for $s>s_2^*$}
\end{cases}
\end{equation}
where $s_2^*$ is given by $\lambda_0^*(s_2^*)=\tilde\lambda_+$. Here
$s_2^* < s_1^*$, as $\tilde{\lambda}_+ > \tilde\lambda_-$.  Finally,
for (P4), in the limit $\delta\to0$, we get
\begin{equation}
I(s)=
\begin{cases}
\mu_0(\tilde\lambda_+) + \tilde\lambda_+ s &\text{for $s<s_2^*$}\\
I_0(s)  &\text{for $s_2^*<s<s_1^*$}\\
\mu_0(\tilde\lambda_-) + \tilde\lambda_- s &\text{for $s>s_1^*$}
\end{cases}
\end{equation}
Note that in all the above three cases, at the points $s_1^*$ and
$s_2^*$, the LDF $I(s)$ generically has second order discontinuities
--- both $I(s)$ and its first derivative $I'(s)$ are continuous,
whereas the second derivative $I''(s)$ is discontinuous, at these
points. Although, similar discontinuities of the LDF have
also been found earlier in the context of work
fluctuations~\cite{sabhapandit2011, sabhapandit2012, pal2013}, the
origins are quite different. In the present case, they originate from
the singularities of $\mu(\lambda)$, while in~\cite{sabhapandit2011,
sabhapandit2012, pal2013}, they originated from the singularity of
$g(\lambda)$. As explained above, in the present analysis, in the
limit $\delta\to 0$, the singularities of $g(\lambda)$ do not play
any role.

It is now straightforward to obtain the asymmetry function $f(s)$,
piecewise, from the above expressions of $I(s)$.  We find that usually
$f(s)=s$ for small $s$, except for $s_2^* >0$ where one has
$f(s)=2\tilde\lambda_+s$ for small $s$. For large $s$, for the
possibilities (P2)--(P4), $f(s)$ differs significantly from the
small-$s$ behavior.  From the second order discontinuities of $I(s)$
at the points $s_{1,2}^*$, it follows that $f(s)$ also exhibits second order discontinuities at these points. The asymptotic
expression of $f(s)$, as $s\to\infty$, is given by
\begin{equation}
f(s)=
\begin{cases}
s &\text{for
 (P1)}\\[1mm]
\bigl[\mu_0(\tilde\lambda_-) -\mu_0(\lambda_+)\bigr]
 + [\tilde\lambda_- +\lambda_+]s &\text{for
(P2)}\\[1mm]
\bigl[\mu_0(\lambda_-) -\mu_0(\tilde\lambda_+)\bigr]
+ [\tilde\lambda_+ +\lambda_-]s &\text{for
(P3)}\\[1mm]
\bigl[\mu_0(\tilde\lambda_-) -\mu_0(\tilde\lambda_+)\bigr] +
[\tilde\lambda_+ +\tilde\lambda_-]s &\text{for
 (P4)}\\[1mm]
\end{cases}
\label{f-function}
\end{equation}
and $f(-s)=-f(s)$, where (P1)--(P4) represent the four cases given
in \eref{possibilities}.  Thus if the analytic region of
$\mu(\lambda)$ is bounded by a pair of branch points, such that in the
limit $\delta\to 0$, at least one of the limiting branch points
differs from that of the uncoupled case ($\delta=0$), then for large
$s$, the slope of the asymmetry function differs from unity.  This
prominent deviation is indeed an indication of coupling to an external
system. \Eref{f-function} is our main result, which provides new FT
for the entropy production of a partial system in the weak coupling
limit.  In the following, we demonstrate that the above mechanism can
indeed be realized in real systems by subjecting it to an external
stochastic forcing.

\textbf{The model (coupled Brownian motion).} --
We consider a system of two Brownian particles (denoted by A and B)
coupled to each other by a harmonic potential $U(y)=k y^2/2$, where
$y$ is the separation between them and $k$ is the spring constant.
For simplicity, we set the masses of both the particles to be equal to
$m$. The Hamiltonian of the coupled system is thus given by
\begin{math}
H=(m/2) (v_\mathrm{A}^2 + v_\mathrm{B}^2) + U(y), 
\end{math}
where $v_\mathrm{A}$ and $v_\mathrm{B}$ are the velocities of the
particles A and B respectively.  The whole system is in contact with a
heat bath at a temperature $T$.  We apply an external Gaussian
stochastic force $f_\mathrm{A}(t)$, with mean zero and correlator
$\langle f_\mathrm{A} (t)
f_\mathrm{A}(t') \rangle=2\bar{f}^2\delta(t-t')$, on the particle
A. It is possible that the particle B also experiences an external
force $f_\mathrm{B}(t)$ (which is assumed to be Gaussian with mean
zero), that may be either correlated or independent to the force
applied on the particle A. Therefore, consider two opposite
situations: i) where $f_\mathrm{B}(t)$ is independent of
$f_\mathrm{A}(t)$ with $\langle f_\mathrm{B} (t)
f_\mathrm{B}(t') \rangle = 2\alpha^2 \bar{f}^2 \delta(t-t')$ and ii)
where they are completely correlated to each other, with
$f_\mathrm{B}(t)=\alpha f_\mathrm{A}(t)$.  In addition to $\alpha$, we
introduce two other dimensionless parameters, $\delta=2km/\gamma^2$
and $\theta=\bar{f}^2/(\gamma T)$, where $\gamma$ is the friction
coefficient. The dynamics of the system is described by the
coupled Langevin equations
\begin{equation}
\begin{split}
\dot{y}&=v_\mathrm{A}- v_\mathrm{B}, 
\\
m \dot{v}_\mathrm{A} (t) &= -\gamma v_{\mathrm A} (t)
  - k y(t) +\eta_\mathrm{A} (t) +f_\mathrm{A}(t), \\
m \dot{v}_\mathrm{B} (t) &= -\gamma v_\mathrm{B} (t)  + k y(t) 
+\eta_\mathrm{B}
(t) 
+f_\mathrm{B}(t),
\end{split}
\label{LE}
\end{equation}
where $\eta_\mathrm{A}(t)$ and $\eta_\mathrm{B}(t)$ are Gaussian white
noise due to the thermal bath acting on the particles A and B
respectively. The mean $\langle \eta_\mathrm{A}(t)\rangle
=\langle \eta_\mathrm{B}(t)\rangle =0$ and the the correlations
$\langle \eta_\mathrm{A}(t) \eta_\mathrm{A}(t')\rangle
=\langle \eta_\mathrm{B}(t) \eta_\mathrm{B}(t')\rangle=2\gamma
T \delta(t-t')$ whereas $\eta_\mathrm{A}$ and $\eta_\mathrm{B}$ are
independent of each other as well as independent of the external
stochastic Gaussian forces $f_\mathrm{A}$ and $f_\mathrm{B}$.

\textbf{Partial entropy production (definition).} --  The heat transfer
from the bath to particle A in a time duration $t$ is given
by~\cite{sekimoto2010}, $Q_\mathrm{A}=\int_0^t [\eta_\mathrm{A}
(t')-\gamma v_\mathrm{A}(t')]v_\mathrm{A}(t')\, dt'$, where
$\eta_\mathrm{A}(t')$ is the Gaussian white noise acting on particle A
due to the thermal bath.  Consequently,
$\Sm^\mathrm{A}=-Q_\mathrm{A}/T$ is the entropy production in the
medium due to particle $A$. In the steady state, the change in the
system entropy for particle A is given by~\cite{seifert2005},
$\Ss^\mathrm{A}=\ln \bigl[P_\mathrm{A}^\mathrm{ss}
(v_{\mathrm{A}}(0))/P_\mathrm{A}^\mathrm{ss} (v_\mathrm{A}(t))\bigr]
$, where $P_\mathrm{A}^\mathrm{ss} (v_\mathrm{A})$ is the steady state
distribution of the velocity $v_\mathrm{A}$. The PDF of the total
entropy production, $\St^\mathrm{A}=\Sm^\mathrm{A}+\Ss^\mathrm{A}$,
for the partial system A, for large $t$, satisfies \eref{PDF}, with
$\tau=\tau_\gamma \equiv m/\gamma$ being the viscous relaxation time.

\textbf{Apparent entropy production (definition).} -- While
$\Sm^\mathrm{A}$ considered above is the true partial entropy
production in the medium due to the system A, an experimental observer
without any knowledge about the coupling with system B would model
system A with $k=0$. This gives the entropy production of the medium due
to A in terms of the experimentally obtainable stochastic trajectories
as
\begin{math}
\tSm=W -\frac{m}{2T}\left[v_\mathrm{A}^2(t)
-v_\mathrm{A}^2(0) \right],
\end{math}
where $W=\frac{1}{T}\int_0^t f_\mathrm{A}(t') v_\mathrm{A}(t')\, dt'$.
Similarly, the change in the system entropy would be defined as
$\tSs=\ln \bigl[\tilde{P}_\mathrm{A}^\mathrm{ss}
(v_{\mathrm{A}}(0))/\tilde{P}_\mathrm{A}^\mathrm{ss}
(v_\mathrm{A}(t))\bigr] $, where $\tilde{P}_\mathrm{A}^\mathrm{ss}
(v_\mathrm{A})$ is the steady state distribution of the velocity
$v_\mathrm{A}$ obtained for $k=0$. By adding both parts, we call
$\tSt=\tSm+\tSs$ as apparent entropy production due to system
A. However, to compute this apparent entropy production, we use the
Langevin equations for the full system A and B with $k\not=0$, as in
reality, there is a non-zero coupling.  Evidently, the two definitions
of the entropy coincides for $\delta=0$.

\begin{figure}
\includegraphics[width=.95\hsize]{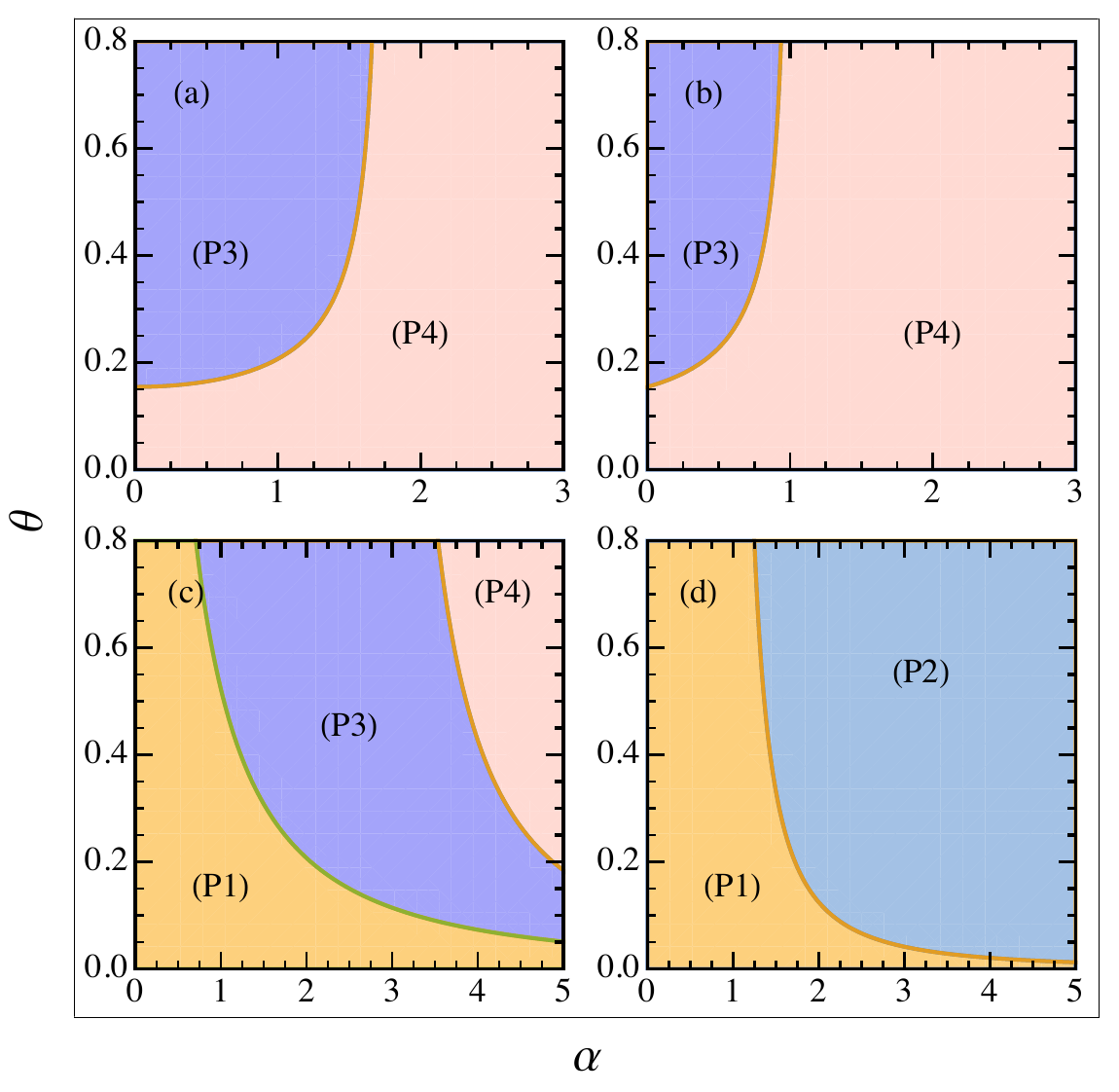}
\caption{\label{RP} (Color online)
The regions in ($\alpha,\theta$) space where the possibilities (P1),
(P2), (P3) and (P4) given in \eref{possibilities} can be realized in
the limit $\delta\to 0$ for: (a) the partial entropy production of
system A and the first choice of $f_\mathrm{B}$, (b) the partial
entropy production of system A and the second choice of
$f_\mathrm{B}$, (c) the apparent entropy production for the first
choice of $f_\mathrm{B}$, and (d) the apparent entropy production for
the second choice of $f_\mathrm{B}$.  }
\end{figure}

\begin{figure*}
\includegraphics[width=.93\hsize]{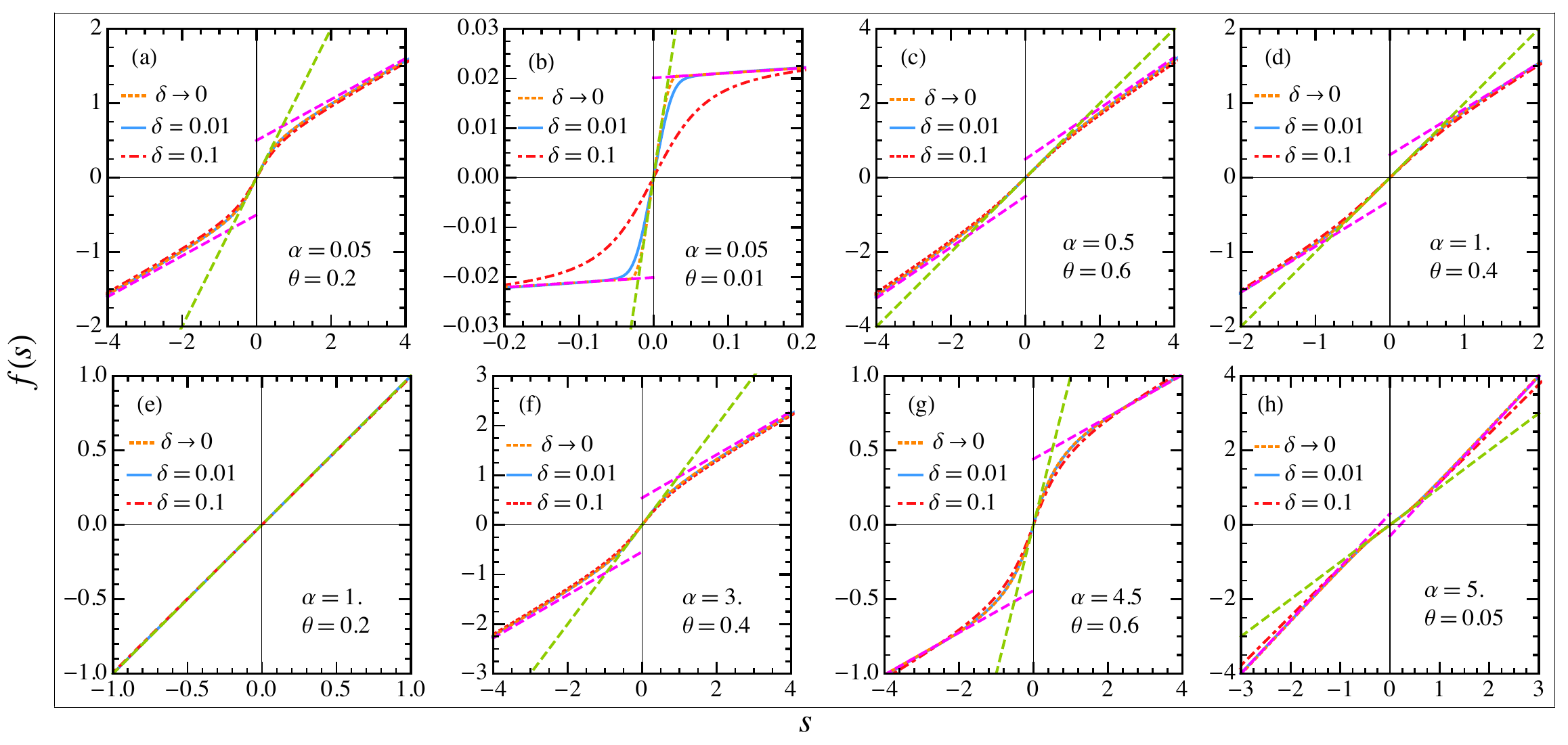}
\caption{\label{SF}  (Color online)
The asymmetry functions corresponding to different regions of the
parameter space of \fref{RP}: (a) and (b) for regions (P3) and (P4)
respectively in \fref{RP}~(a).  (c) and (d) for the regions (P3) and
(P4) respectively in \fref{RP}~(b). (e) for regions (P1)
of \fref{RP}~(c) as well as \fref{RP}~{d}.  (f) and (g) for the
regions (P3) and (P4) respectively in \fref{RP}~(c).  (h) for region
(P2) in \fref{RP}~(d).  In all the figures the green dashed lines
through the origin plot the function $f(s)=s$ and the magenta dashed
lines plot the asymptotic predictions of $f(s)$ for large $s$, given
by \eref{f-function} --- the points where these two lines meet do not
have any significant.  The orange dotted lines (marked by $\delta\to
0$) plot the limiting expressions of $f(s)$, obtained from the
expressions of $I(s)$ given in the text.}
\end{figure*}


\textbf{Methods for computation of $\mu(\lambda)$ and $g(\lambda)$.} --
It is convenient to use the Fourier transforms in the time domain
$(0,t)$.  Since the equations \erefss{LE} are linear, the Fourier
transforms $(\tilde{y},\tilde{v}_\mathrm{A},
\tilde{v}_\mathrm{B})$, depend linearly on $(\tilde{f}_\mathrm{A},
\tilde{f}_\mathrm{B}, \tilde{\eta}_\mathrm{A}, \tilde{\eta}_\mathrm{B})$,
Consequently, $\Sm^\mathrm{A}$ and $W$ are quadratic in
$(\tilde{f}_\mathrm{A}, \tilde{f}_\mathrm{B}, \tilde{\eta}_\mathrm{A}, 
\tilde{\eta}_\mathrm{B})$. Since, stochastic forces and the thermal
noises are uncorrelated in time, their Fourier transforms for any
frequency $\omega$ are correlated to only negative frequency
$-\omega$. Moreover, since $(f_\mathrm{A},
f_\mathrm{B}, \eta_\mathrm{A}, \eta_\mathrm{B})$ are real variables,
their Fourier transforms at a negative frequency $-\omega$, are equal
to the complex conjugate of the corresponding Fourier transforms with
positive frequency $\omega$, i.e.,
$\tilde{f}_\mathrm{A}(-\omega)=\tilde{f}^*_\mathrm{A}(\omega)$ and so
on.  Therefore, using the Gaussian distribution of
$(\tilde{f}_\mathrm{A},
\tilde{f}_\mathrm{B}, \tilde{\eta}_\mathrm{A}, \tilde{\eta}_\mathrm{B})$
independently for each frequency, we can express the generating
functions $\langle \exp(-\lambda \St^\mathrm{A})\rangle$ and
$\langle \exp(-\lambda \tSt)\rangle$ as infinite product of
independent Gaussian integrals for each frequency.  Finally, in the
large-$t$ limit, by considering the frequency to be continuous, and
integrating over the final phase space variables and averaging over
the initial phase space variables, we obtain (see \cite{Kundu2011} for
details on a similar derivation) the generating function similar
to \eref{Z} with $\tau=m/\gamma$. Since, system entropy productions
depend only on the initial and final velocities, and not on the full
trajectory, they do not contribute to $\mu(\lambda)$, but contribute
only to the prefactor $g(\lambda)$. The calculations are carried
out\footnote{\label{note2}Details will be published elsewhere.} separately for the four cases: i.e., the two definitions
of the entropy and the two choices of the stochastic force
$f_\mathrm{B}$. In all the cases, $\mu(\lambda)$ has the integral form
\begin{equation}
\mu(\lambda)=-\dfrac{1}{4\pi
}\int_{-\infty}^{\infty}du\, \ln{\left[1+\dfrac{h(u,\lambda)}{q(u)} \right]}
,
\label{mu}
\end{equation}
where the functions $h(u,\lambda)$ and $q(u)$ are different for each
case. The expressions of $g(\lambda)$, for all the cases are quite
involved and not very illuminating. Fortunately, we are interested in
the limit of $\delta\to 0$ (i.e., $k\to 0$), and for that, as
discussed above, we only need $g_0(\lambda)$.  This is given by a
simple expression (see footnote \textsuperscript{\ref{note2}}), $
g_0(\lambda)=2 \sqrt{\nu(\lambda)}/[1+\nu(\lambda)]$ with
$\nu(\lambda)=
\sqrt{1+4\theta\lambda(1-\lambda)}
$.

\textbf{Partial entropy production (results).} --
For the uncoupled case ($\delta=0$), we find that (see footnote \textsuperscript{\ref{note2}})
$\mu_0(\lambda)=\frac{1}{2}\bigl[1-\nu(\lambda)\bigr]$ has branch
point singularities at
$\lambda_\pm=\frac{1}{2}\bigl[1\pm\sqrt{1+\theta^{-1}}\bigr]$ and is
analytic in the region bounded by these branch points. We also find
that $I_0(s)$ obeys \eref{LDF symmetry}. Now, in the presence of a
non-zero coupling $\delta$, we find that
$\lambda_+^{(\delta)}=\tilde\lambda_+ =1$ for all $0<\delta <1$,
including the limit $\delta\to 0$. On the other hand, the limiting
behavior of $\lambda_-^{(\delta)}$ depends on the parameters
$(\alpha,\theta)$ as shown in \fref{RP}, where the phase boundaries
can be calculated exactly (see footnote~\textsuperscript{\ref{note2}}). In the (P3) regions in figs.~\ref{RP}(a) and (b), we get
$\lambda_-^{(\delta)}\to\lambda_-$ as $\delta\to 0$ whereas in the
(P4) regions in figs.~\ref{RP}(a) and (b), we get
$\lambda_-^{(\delta)}=\tilde\lambda_-$ for all $0<\delta <1$. Here
$\tilde\lambda_- = -[1+(1+\alpha^2)\theta]^{-1}$ for the choice i) of
$f_\mathrm{B}(t)$ and $\tilde\lambda_- = -[1+(1+\alpha)^2\theta]^{-1}$
for the choice ii).  Therefore, in the limit $\delta\to 0$, while
$f(s)\approx s$ for small $s$, it is given by \eref{f-function} for
large $s$.

\textbf{Apparent entropy production (results).} --
 In the presence of a non-zero $\delta$, for the choice i) of
$f_\mathrm{B}(t)$, we find that the possibilities (P1), (P3) and (P4)
can be realized (\fref{RP}(c)) depending on the values of the driving
parameters (see footnote~\textsuperscript{\ref{note2}}), and $\tilde\lambda_\pm$ given by
$\tilde\lambda_\pm= \bigl\{\theta\pm\sqrt{\theta
[2+(1+\alpha^{2})\theta]}\bigr\}\big/\bigl\{\theta(2+\alpha^{2}\theta)\bigr\}$.
On the other hand, for the choice ii) of $f_\mathrm{B}(t)$, the
possibilities (P1) and (P2) can be realized (\fref{RP}~(d) and
$\tilde\lambda_-=\bigl\{ (1+\alpha) \theta -\sqrt{\theta
[2+(1+\alpha)^{2}\theta]}\bigr\}\big/(2\theta) $.

\textbf{Numerical comparison.} --
In \fref{SF} (also see footnote~\textsuperscript{\ref{note2}}), we compare the theoretical
prediction of $f(s)$ for large $s$ in the limit $\delta\to 0$, given
by \eref{f-function}, against exact results obtained by numerically
inverting $\mu(\lambda)$ for small $\delta$, for both the choices of
the driving force $f_\mathrm{B}(t)$ and also for both the
considerations of the entropy production.  We find that, as $\delta$
decreases, the numerical curves converge to the limiting (limit
$\delta\to 0$) expressions of $f(s)$ (see footnote~\textsuperscript{\ref{note2}}). Moreover,
for large $s$, they converge to the asymptotic expressions given
in \eref{f-function}. We also refer to footnote~\textsuperscript{\ref{note2}} for comparisons of
the PDF and LDF with numerics.

\textbf{System in a trap.} --
Finally, we ask whether the effect of the external weak coupling can
be nullified.  Indeed, we find that (see footnote~\textsuperscript{\ref{note2}}), when the system is
placed in a harmonic trap, we always get
$\lambda_\pm^{(\delta)} \to \lambda_\pm$ as $\delta\to 0$. Thus, in
this case, the relation $f(s)=s$ is always satisfied in the limit
$\delta\to0$. This suggests that weak coupling cannot affect the FT in
the presence of a trap, and hence, this provides a way to neutralize
the influence of such coupling.

\textbf{Concluding remarks.} --
We have found a new mechanism by which the FT of the entropy
production can be violated in the presence of a coupling to an
external system, even in the limit of the coupling strength going to
zero.  In other words, we have provided a new FT for the entropy
production of a partial system, in the presence of weak coupling,
driven by external random forces.  Conversely, our finding gives a way
to find out if a particular stochastic process of interest is coupled
to any other hidden stochastic systems.  Thus, it provides a new
application of FT that can be applied to a wide variety of small
systems.

Our result may look quite surprising at first glance, as it goes
against our naive intuition that the effect of coupling should
disappear in the limit of interaction strength going to zero, as in
the regular perturbation problems. However, it should be emphasized
that, here the limit of coupling strength going to zero is a singular
perturbation, which is very different from the case of coupling
strength equal to zero.  For example, in the case of the two coupled
Brownian particles A and B (without the external forces), the coupling
introduces a timescale $\tau_k=\gamma/k$, beyond which the the
separation between the particles relaxes to the equilibrium with
$\langle y^2 \rangle_\mathrm{eq} = D\tau_k$, whereas for
$\tau_\gamma\ll t\ll \tau_k$ we have $\langle\Delta y^2\rangle = 4Dt$,
where $D$ is the diffusion constant.  Therefore, while initially the
particle A (or B) behaves like a free particle with $\langle \Delta
x_\mathrm{A}^2 \rangle = 2Dt$ for $\tau_\gamma\ll t \ll \tau_k$,
beyond the timescale $\tau_k$, it diffuses as the center of mass with
$\langle \Delta x_\mathrm{A}^2 \rangle = (D\tau_k)/4+ Dt$ for
$t\gg \tau_k$.  This example indirectly demonstrates that
it is possible to observe the effect of weak coupling if one looks at
a time beyond the timescale introduced by the
coupling. Note that in this Letter, we have already taken
the large time limit before taking the $\delta\to 0$ limit. However,
it does not directly explain our results, as we have found that in the
regions (P1) of figs.~\ref{RP}(c) and (d), the FT is
satisfied. A clear understanding of how different time
scales can lead to the singular limit, remains an open problem. It
would be interesting to demonstrate the crossover from the validity of
the FT at small time to the singular limit at long time, through a
model that can be exactly solved for all time.

 The harmonic trap (of stiffness $k_0$) introduces a timescale
$\tau_{k_0}=\gamma/k_0$ beyond which the system relaxes to the steady
state. The weak coupling changes this timescale only by a small amount
$\tau_k = \tau_{k_0} [1+O(k/k_0)]$. Perhaps, for this reason, in the
presence of a trap, the FT is always satisfied in the limit of
coupling strength going to zero.


\begin{thebibliography}{10}
\expandafter\ifx\csname url\endcsname\relax\def\url#1{\texttt{#1}}\fi

\bibitem{sekimoto2010}
\Name{Sekimoto K.} \Book{Stochastic Energetics} Vol. 799 of \emph{Lect. Notes.
  Phys.} (Springer, New York) 2010.

\bibitem{jarzynski2011}
\Name{Jarzynski C.} \REVIEW{Ann. Rev. Cond. Mat. Phys.}{2}{2011}{329}.


\bibitem{sevick2008}
\Name{Sevick E.~M., Prabhakar R., Williams S.~R. \and Searles D.~J.}
  \REVIEW{Ann. Rev. Phys. Chem.}{59}{2008}{603}.


\bibitem{Campisi:2011}
\Name{Campisi M., H{\"a}nggi P. \and Talkner P.} \REVIEW{Rev. Mod. Phys.}{83}{2011}{771}.


\bibitem{seifert2012}
\Name{Seifert U.} \REVIEW{Rep. Prog. Phys.}{75}{2012}{126001}.


\bibitem{Van-den-Broeck:2015}
\Name{Van~den Broeck C. \and Esposito M.} \REVIEW{Physica A}{418}{2015}{6}.


\bibitem{Esposito:2009}
\Name{Esposito M., Harbola U. \and Mukamel S.} \REVIEW{Rev. Mod. Phys.}{81}{2009}{1665}.


\bibitem{bustamante2005}
\Name{Bustamante C., Liphardt J. \and Ritort F.} \REVIEW{Phys. Today}{58}{2005} {43}.


\bibitem{verley2014}
\Name{Verley G., Esposito M., Willaert T. \and {V}an~den Broeck C.}
  \REVIEW{Nature Comm.}{5}{2014}{4721}.


\bibitem{martinez2015}
\Name{Martinez I.~A., Roldan E., Dinis L., Petrov D., Parrondo J. M.~R. \and
  Rica R.~A.} \REVIEW{Nat. Phys.}{12}{2016}{67}.


\bibitem{blickle2011}
\Name{Blickle V. \and Bechinger C.} \REVIEW{Nat. Phys.}{8}{2011}{143}.


\bibitem{zon2004a}
\Name{van Zon R., Ciliberto S. \and Cohen E. G.~D.} \REVIEW{Phys. Rev. Lett.}{92}{2004}{130601}.


\bibitem{tietz2006}
\Name{Tietz C., Schuler S., Speck T., Seifert U. \and Wrachtrup J.}
  \REVIEW{Phys. Rev. Lett.}{97}{2006}{050602}.


\bibitem{wang2002}
\Name{Wang G.~M., Sevick E.~M., Mittag E., Searles D.~J. \and Evans D.~J.}
  \REVIEW{Phys. Rev. Lett.}{89}{2002}{050601}.

\bibitem{Carberry:2004aa}
\Name{Carberry D.~M., Reid J.~C., Wang G.~M., Sevick E.~M., Searles D.~J. \and
  Evans D.~J.} \REVIEW{Phys. Rev. Lett.}{92}{2004}{140601}.


\bibitem{liphardt2002}
\Name{Liphardt J., Dumont S., Smith S.~B., Tinoco I. \and Bustamante C.}
  \REVIEW{Science}{296}{2002}{1832}.


\bibitem{collin2005}
\Name{Collin D., Ritort F., Jarzynski C., Smith S.~B., Tinoco I. \and
  Bustamante C.} \REVIEW{Nature}{437}{2005}{231}.


\bibitem{Koski:2013aa}
\Name{Koski J.~V., Sagawa T., Saira O.-P., Yoon Y., Kutvonen A., Solinas P.,
  Mottonen M., Ala-Nissila T. \and Pekola J.~P.} \REVIEW{Nat. Phys.
  }{9}{2013}{644}.


\bibitem{Ciliberto:2013aa}
\Name{Ciliberto S., Gomez-Solano R. \and Petrosyan A.} \REVIEW{Ann. Rev. Cond.
  Matt. Phys.}{4}{2013}{235}.


\bibitem{evans1993}
\Name{Evans D.~J., Cohen E. G.~D. \and Morriss G.~P.} \REVIEW{Phys. Rev. Lett.}{71}{1993}{2401}.

\bibitem{gallavotti1995}
\Name{Gallavotti G. \and Cohen E. G.~D.} \REVIEW{Phys. Rev. Lett.}{74}{1995}{2694}.

\bibitem{kurchan1998}
\Name{Kurchan J.} \REVIEW{J. Phys. A: Math. Gen.}{31}{1998}{3719}.


\bibitem{lebowitz1999}
\Name{Lebowitz J.~L. \and Spohn H.} \REVIEW{J. Stat. Phys.}{95}{1999}{333}.


\bibitem{seifert2005}
\Name{Seifert U.} \REVIEW{Phys. Rev. Lett.}{95}{2005}{040602}.


\bibitem{jarzynski1997}
\Name{Jarzynski C.} \REVIEW{Phys. Rev. Lett.}{78}{1997}{2690}.

\bibitem{crooks1999}
\Name{Crooks G.~E.} \REVIEW{Phys. Rev. E}{60}{1999}{2721}.

\bibitem{zon2003}
\Name{van Zon R. \and Cohen E. G.~D.} \REVIEW{Phys. Rev. Lett.}{91}{2003}{110601}.

\bibitem{Saito:2007}
\Name{Saito K. \and Dhar A.} \REVIEW{Phys. Rev. Lett.}{99}{2007}{180601}.


\bibitem{Speck:2005}
\Name{Speck T. \and Seifert U.} \REVIEW{J. Phys. A: Math. Gen.}{38}{2005}{L581}.


\bibitem{Jarzynski:2004}
\Name{Jarzynski C. \and W{\'o}jcik D.~K.} \REVIEW{Phys. Rev. Lett.}{92}{2004}{230602}.


\bibitem{Noh:2012}
\Name{Noh J.~D. \and Park J.-M.} \REVIEW{Phys. Rev. Lett.}{108}{2012}{240603}.


\bibitem{Mehl:2012fw}
\Name{Mehl J., Lander B., Bechinger C., Blickle V. \and Seifert U.}
  \REVIEW{Phys. Rev. Lett.}{108}{2012}{220601}.


\bibitem{Pietzonka:2014bf}
\Name{Pietzonka P., Zimmermann E. \and Seifert U.} \REVIEW{Europhys. Lett.}{107}{2014}{20002}.


\bibitem{Borrelli:2015eq}
\Name{Borrelli M., Koski J.~V., Maniscalco S. \and Pekola J.~P.} \REVIEW{Phys.
  Rev. E }{91}{2015}{012145}.


\bibitem{Ribezzi-Crivellari:2014}
\Name{Ribezzi-Crivellari M. \and Ritort F.} \REVIEW{Proc. Nat. Acad. Sci. U.S.A}{111}{2014}{E3386}.


\bibitem{Chun2015}
\Name{Chun H.-M. \and Noh J.~D.} \REVIEW{Phys. Rev. E}{91}{2015}{052128}.


\bibitem{Rahav:2007aa}
\Name{Rahav S. \and Jarzynski C.} \REVIEW{J. Stat. Mech.}{2007}{2007}{P09012}.


\bibitem{Puglisi:2010aa}
\Name{Puglisi A., Pigolotti S., Rondoni L. \and Vulpiani A.} \REVIEW{J.
  Stat. Mech.}{2010}{2010}{P05015}.


\bibitem{Shiraishi:2015aa}
\Name{Shiraishi N. \and Sagawa T.} \REVIEW{Phys. Rev. E
  }{91}{2015}{012130}.


\bibitem{Talkner:2009aa}
\Name{Talkner P., Campisi M. \and H{\"a}nggi P.} \REVIEW{J.  Stat.
  Mech.}{2009}{2009}{P02025}.


\bibitem{touchette2009}
\Name{Touchette H.} \REVIEW{Phys. Rep.}{478}{2009}{1}.


\bibitem{sabhapandit2011}
\Name{Sabhapandit S.} \REVIEW{Europhys. Lett.}{96}{2011}{20005}.


\bibitem{sabhapandit2012}
\Name{Sabhapandit S.} \REVIEW{Phys. Rev. E}{85}{2012}{021108}.


\bibitem{pal2013}
\Name{Pal A. \and Sabhapandit S.} \REVIEW{Phys. Rev. E}{87}{2013}{022138}.


\bibitem{Kundu2011} 
\Name{Kundu A., Sabhapandit S., \and Dhar
 A.} \REVIEW{J. Stat. Mech.}{}{2011}{P03007}.

\end{thebibliography}


\end{document}